\documentclass[12pt]{revtex4}
\usepackage{amssymb,epsf}
\usepackage{latexsym}
\begin{document}

\title{Charged rotating dilaton black branes in
AdS universe}
\author{A. Sheykhi$^{1,2}$\footnote{sheykhi@mail.uk.ac.ir} and S.H.
Hendi$^{3}$\footnote{hendi@mail.yu.ac.ir}}

\address{$^1$ Department of Physics, Shahid Bahonar University, P.O. Box 76175, Kerman, Iran\\
         $^2$ Research Institute for Astrophysics and Astronomy of Maragha (RIAAM), Maragha, Iran\\
         $^3$ Department of Physics, College of Science, Yasouj University, Yasouj 75914, Iran}

\begin{abstract}
We present the metric for the $(n+1)$-dimensional charged rotating
dilaton black branes with cylindrical or toroidal horizons in the
background of anti-de Sitter spacetime. We find the suitable
counterterm which removes the divergences of the action in the
presence of the dilaton potential in all higher dimensions. We
plot the  Penrose diagrams of the spacetime and reveal that the
spacetime geometry  crucially modifies in the presence of the
dilaton field. The conserved and thermodynamic quantities of the
black branes are also computed.
\end{abstract}

\maketitle

\section{Introduction\label{Intr}}

The motivation idea for studying higher dimensional black holes
with a negative cosmological constant arises from the
correspondence between the gravitating fields in an anti-de Sitter
(AdS) spacetime and conformal field theory living on the boundary
of the AdS spacetime \cite{Witt1}. It was argued that the
thermodynamics of black holes in AdS spaces can be identified with
that of a certain dual CFT in the high temperature limit
\cite{Witt2}. Having the AdS/CFT correspondence idea at hand, one
can gain some insights into thermodynamic properties and phase
structures of strong 't Hooft coupling CFTs by studying
thermodynamics of AdS black holes. According to the AdS/CFT
correspondence, the rotating black holes in AdS space are dual to
certain CFTs in a rotating space \cite{Haw}, while charged ones
are dual to CFTs with chemical potential \cite{Cham1,Cham2}. The
most general higher dimensional
uncharged rotating black holes in AdS space have been recently found \cite%
{Haw,Gib}. As far as we know, rotating black holes for the Maxwell field
minimally coupled to Einstein gravity in higher dimensions, do not exist in
a closed form and one has to rely on perturbative or numerical methods to
construct them in the background of asymptotically flat \cite{kunz1,Aliev2}
and AdS \cite{kunz2} spacetimes. There has also been recent interest in
constructing the analogous charged rotating solutions in the framework of
gauged supergravity in various dimensions \cite{Cvetic0,Cvetic1,Cvetic2}.

There has been a renewed interest in studying scalar coupled solutions of
general relativity ever since new black hole solutions have been found in
the context of string theory. The low energy effective action of string
theory contains two massless scalars namely dilaton and axion. The dilaton
field couples in a nontrivial way to other fields such as gauge fields and
results into interesting solutions for the background spacetime \cite%
{CDB1,CDB2,Hor2,Hor3}. These solutions \cite{CDB1,CDB2,Hor2,Hor3},
however, are all asymptotically flat. The presence of
Liouville-type dilaton potential, which is regarded as the
generalization of the cosmological constant, changes the
asymptotic behavior of the solutions to be neither asymptotically
flat nor (A)dS. While in the presence of one or two Liouville-type
potential, black holes/branes which are neither asymptotically
flat nor (A)dS have been explored in
\cite{MW,CHM,Cai,Clem,Deh1,Deh2,Sheykhi0,Sheykhi1,DPH,DHSR},
magnetic dilaton solutions coupled to nonlinear electrodynamics
have also been investigated \cite{DSH}. Although these kind of
solutions may shed some light on the possible extensions of
AdS/CFT correspondence, they are physically less interesting due
to their unusual asymptotic behavior. Recently, the dilaton
potential leading to (A)dS-like solutions of dilaton gravity has
been found \cite{Gao1,Gao2,Gao3}. It was shown that the
cosmological constant is coupled to the dilaton in a very
nontrivial way. With an appropriate combination of three
Liouville-type dilaton potentials, a class of static dilaton black
hole solutions in (A)dS spaces has been obtained by using a
coordinates transformation which recast the solution in the
schwarzschild coordinates system \cite{Gao1,Gao2}. Such potential
may arise from the compactification of a higher dimensional
supergravity model \cite{Gid} which originates from the low energy
limit of a background string theory. More recently, one of us has
constructed a class of magnetic rotating solutions in
four-dimensional Einstein-Maxwell-dilaton
gravity with Liouville-type potential in the background of AdS spaces \cite%
{Sheykhi2}. Although these solutions are not black holes and represent
spacetimes with conic singularities, asymptotically AdS charged rotating
black string solutions in four-dimensional Einstein-Maxwell-dilaton gravity
has also been constructed \cite{Sheykhi3}. So far, exact higher dimensional
charged rotating dilaton black hole/brane solutions for an arbitrary
dilaton-electromagnetic coupling constant in the background of AdS spacetime
have not been constructed. In this paper we intend to construct exact,
charged rotating dilaton black branes with cylindrical or toroidal horizons
in higher dimensional AdS spacetimes and investigate their properties.

This paper is outlined as follows. Section \ref{Field} is devoted
to a brief review of the field equations and the general formalism
of calculating the conserved quantities. We shall also present the
suitable counterterm which removes the divergences of the action
in the presence of the dilaton potential. In section
\ref{Charged}, we construct the $(n+1)$-dimensional charged
rotating dilaton black branes with a complete set of rotation
parameters and investigate their properties. We also obtain the
conserved and thermodynamical quantities of the
$(n+1)$-dimensional black brane solutions. We finish with
conclusion in the last section.

\section{Basic Equations and counterterm method}

\label{Field} We consider the $(n+1)$-dimensional theory in which gravity is
coupled to dilaton and Maxwell field with an action
\begin{eqnarray}
I_{G} &=&-\frac{1}{16\pi }\int_{\mathcal{M}}d^{n+1}x\sqrt{-g}\left( {R}\text{
}-\frac{4}{n-1}(\nabla \Phi )^{2}-V(\Phi )-e^{-4\alpha \Phi /(n-1)}F_{\mu
\nu }F^{\mu \nu }\right)  \nonumber \\
&&-\frac{1}{8\pi }\int_{\partial \mathcal{M}}d^{n}x\sqrt{-\gamma }\Theta
(\gamma ),  \label{Act}
\end{eqnarray}
where ${R}$ is the scalar curvature, $\Phi$ is the dilaton field, $F_{\mu
\nu }=\partial _{\mu }A_{\nu }-\partial _{\nu }A_{\mu }$ is the
electromagnetic field tensor, and $A_{\mu }$ is the electromagnetic
potential. $\alpha $ is an arbitrary constant governing the strength of the
coupling between the dilaton and the Maxwell field. The last term in Eq. (%
\ref{Act}) is the Gibbons-Hawking surface term. It is required for the
variational principle to be well-defined. The factor $\Theta$ represents the
trace of the extrinsic curvature for the boundary ${\partial \mathcal{M}}$
and $\gamma$ is the induced metric on the boundary. While $\alpha=0$
corresponds to the usual Einstein-Maxwell-scalar theory, $\alpha=1$
indicates the dilaton-electromagnetic coupling that appears in the low
energy string action in Einstein's frame. For arbitrary value of $\alpha $
in (A)dS spaces the form of the dilaton potential in arbitrary dimensions is
chosen as \cite{Gao2}
\begin{eqnarray}  \label{V1}
{V}({\Phi})&=&\frac{2\Lambda}{n(n-2+\alpha^2)^{2}} \Big{\{}-\alpha^2\left[%
(n+1)^{2}-(n+1)\alpha^{2}-6(n+1)+\alpha^{2}+9\right] e^{{-4(n-2){\Phi}}/{%
[(n-1)\alpha]}}  \nonumber \\
&& +(n-2)^{2}(n-\alpha^{2}) e^{{4\alpha{\Phi}}/({n-1})}+4%
\alpha^{2}(n-1)(n-2) e^{{-2{\Phi}(n-2-\alpha^{2})}/{[(n-1)\alpha}]}\Big {\}}%
.
\end{eqnarray}
Here $\Lambda $ is the cosmological constant. For later convenience we
redefine $\Lambda=-n(n-1)/2l^2$, where $l$ is the AdS radius of spacetime.
It is clear the cosmological constant is coupled to the dilaton in a very
nontrivial way. This type of the dilaton potential was introduced for the
first time by Gao and Zhang \cite{Gao2}. They derived, by applying a
coordinates transformation which recast the solution in the Schwarzchild
coordinates system, the static dilaton black hole solutions in the
background of (A)dS universe. For this purpose, they required the existence
of the (A)dS dilaton black hole solutions and extracted successfully the
form of the dilaton potential leading to (A)dS-like solutions. They also
argued that this type of derived potential can be obtained when a higher
dimensional theory is compactified to four dimension, including various
supergravity models \cite{Gid}. In the absence of the dilaton field ($\Phi
=0=\alpha$), the potential (\ref{V1}) reduces to ${V}({\Phi})=2\Lambda$, and
the action (\ref{Act}) recovers the action of Einstein-Maxwell gravity with
cosmological constant. The equations of motion can be obtained by varying
the action (\ref{Act}) with respect to the gravitational field $g_{\mu \nu }$%
, the dilaton field $\Phi $ and the gauge field $A_{\mu }$ which yields the
following field equations
\begin{equation}
\mathcal{R}_{\mu \nu }=\frac{4}{n-1}\left( \partial _{\mu }\Phi \partial
_{\nu }\Phi +\frac{1}{4}g_{\mu \nu }V(\Phi )\right) +2e^{{-4\alpha \Phi }/{%
(n-1)}}\left( F_{\mu \eta }F_{\nu }^{\text{ }\eta }-\frac{1}{2(n-1)}g_{\mu
\nu }F_{\lambda \eta }F^{\lambda \eta }\right) ,  \label{FE1}
\end{equation}
\begin{equation}
\nabla ^{2}\Phi =\frac{n-1}{8}\frac{\partial V}{\partial \Phi }-\frac{\alpha
}{2}e^{-{4\alpha \Phi }/{(n-1})}F_{\lambda \eta }F^{\lambda \eta },
\label{FE2}
\end{equation}
\begin{equation}
\partial _{\mu }\left( \sqrt{-g}e^{{-4\alpha \Phi }/({n-1})}F^{\mu \nu
}\right) =0.  \label{FE3}
\end{equation}
The conserved charges of the spacetime can be calculated through the use of
the substraction method of Brown and York \cite{BY}. Such a procedure causes
the resulting physical quantities to depend on the choice of reference
background. For asymptotically AdS solutions, the way that one can calculate
these quantities and obtain finite values for them is through the use of the
counterterm method inspired by AdS/CFT correspondence \cite{Witt1}. In this
paper we deal with the spacetimes with zero curvature boundary, $%
R_{abcd}(\gamma )=0$, and therefore the counterterm for the stress energy
tensor should be proportional to $\gamma ^{ab}$. We find the suitable
counterterm which removes the divergences in the form
\begin{equation}  \label{cont}
I_{ct}=-\frac{1}{8\pi }\int_{\partial \mathcal{M}}d^{n}x\sqrt{-\gamma }%
\left(-\frac{(n-1)(n-2)}{2l}+\frac{\sqrt{-n(n-1)V(\Phi)}}{2} \right).
\end{equation}
In the absence of the dilaton field, ${V}({\Phi})=2\Lambda=-n(n-1)/l^2$, and
Eq. (\ref{cont}) reduces to
\begin{equation}  \label{cont2}
I_{ct}=-\frac{1}{8\pi }\int_{\partial \mathcal{M}}d^{n}x\sqrt{-\gamma }\left(%
\frac{n-1}{l}\right),
\end{equation}
which is the counterterm of the asymptotically AdS spaces. Having the total
finite action $I = I_{G}+I_{\mathrm{ct}}$ at hand, we can use the quasilocal
definition to construct a divergence free stress-energy tensor \cite{BY}.
Thus we write down the finite stress-energy tensor in $(n+1)$-dimensional
Einstein-dilaton gravity with three Liouville-type dilaton potentials (\ref%
{V1}) in the following form
\begin{equation}
T^{ab}=\frac{1}{8\pi }\left[ \Theta ^{ab}-\Theta \gamma ^{ab}+\left(-\frac{%
(n-1)(n-2)}{2l}+\frac{\sqrt{-n(n-1)V(\Phi)}}{2} \right)\gamma ^{ab}\right].
\label{Stres}
\end{equation}
The first two terms in Eq. (\ref{Stres}) are the variation of the action (%
\ref{Act}) with respect to $\gamma _{ab}$, and the last two terms are the
variation of the boundary counterterm (\ref{cont}) with respect to $\gamma
_{ab}$. To compute the conserved charges of the spacetime, one should choose
a spacelike surface $\mathcal{B}$ in $\partial \mathcal{M}$ with metric $%
\sigma _{ij}$, and write the boundary metric in ADM (Arnowitt-Deser-Misner)
form:
\[
\gamma _{ab}dx^{a}dx^{a}=-N^{2}dt^{2}+\sigma _{ij}\left( d\varphi
^{i}+V^{i}dt\right) \left( d\varphi ^{j}+V^{j}dt\right) ,
\]
where the coordinates $\varphi ^{i}$ are the angular variables
parameterizing the hypersurface of constant $r$ around the origin, and $N$
and $V^{i}$ are the lapse and shift functions respectively. When there is a
Killing vector field $\mathcal{\xi }$ on the boundary, then the quasilocal
conserved quantities associated with the stress tensors of Eq. (\ref{Stres})
can be written as
\begin{equation}
Q(\mathcal{\xi )}=\int_{\mathcal{B}}d^{n-1}x \sqrt{\sigma }T_{ab}n^{a}%
\mathcal{\xi }^{b},  \label{charge}
\end{equation}
where $\sigma $ is the determinant of the metric $\sigma _{ij}$, $\mathcal{%
\xi }$ and $n^{a}$ are, respectively, the Killing vector field and the unit
normal vector on the boundary $\mathcal{B}$. For boundaries with timelike ($%
\xi =\partial /\partial t$) and rotational ($\varsigma =\partial /\partial
\varphi $) Killing vector fields, one obtains the quasilocal mass and
angular momentum
\begin{eqnarray}
M &=&\int_{\mathcal{B}}d^{n-1}x \sqrt{\sigma }T_{ab}n^{a}\xi ^{b},
\label{Mastot} \\
J &=&\int_{\mathcal{B}}d^{n-1}x \sqrt{\sigma }T_{ab}n^{a}\varsigma ^{b}.
\label{Angtot}
\end{eqnarray}
provided the surface $\mathcal{B}$ contains the orbits of $\varsigma $.
These quantities are, respectively, the conserved mass and angular momenta
of the system enclosed by the boundary $\mathcal{B}$. Note that they will
both depend on the location of the boundary $\mathcal{B}$ in the spacetime,
although each is independent of the particular choice of foliation $\mathcal{%
B}$ within the surface $\partial \mathcal{M}$.


\section{Charged rotating dilaton black brane}

\label{Charged} \label{field}Our aim here is to construct the $(n+1)$%
-dimensional rotating solutions of the field equations (\ref{FE1})-(\ref{FE3}%
) with $k$ rotation parameters and investigate their properties. The
rotation group in $(n+1)$-dimensions is $SO(n)$ and therefore the number of
independent rotation parameters for a localized object is equal to the
number of Casimir operators, which is $[n/2]\equiv k$, where $[x]$ is the
integer part of $x$. Inspired by \cite{awad}, we take the metric of $(n+1)$%
-dimensional rotating solution with cylindrical or toroidal horizons and $k$
rotation parameters in the form
\begin{eqnarray}
ds^{2} &=&-U(r)\left( \Xi dt-{{\sum_{i=1}^{k}}}a_{i}d\phi _{i}\right) ^{2}+%
\frac{r^{2}}{l^{4}}R^{2}(r){{\sum_{i=1}^{k}}}\left( a_{i}dt-\Xi l^{2}d\phi
_{i}\right) ^{2}  \nonumber  \label{metric} \\
&&-\frac{r^{2}}{l^{2}}R^{2}(r){\sum_{i<j}^{k}}(a_{i}d\phi _{j}-a_{j}d\phi
_{i})^{2}+\frac{dr^{2}}{W(r)}+\frac{r^{2}}{l^{2}}R^{2}(r)dX^{2},  \nonumber
\\
\Xi ^{2} &=&1+\sum_{i=1}^{k}\frac{a_{i}^{2}}{l^{2}},  \label{Met3}
\end{eqnarray}%
where $a_{i}$'s are $k$ rotation parameters. The functions $U(r)$, $W(r)$
and $R(r)$ should be determined and $l$ has the dimension of length which is
related to the cosmological constant $\Lambda $ for the case of
Liouville-type potential with constant $\Phi $. The angular coordinates are
in the range $0\leq \phi _{i}\leq 2\pi $ and $dX^{2}$ is the Euclidean
metric on the $(n-k-1)$-dimensional submanifold with volume $\Sigma _{n-k-1}$%
. The Maxwell equation (\ref{FE3}) can be integrated immediately to give
\begin{eqnarray}
F_{tr} &=&\sqrt{\frac{U(r)}{W(r)}}\frac{q\Xi e^{{4\alpha \Phi }/{(n-1)}}}{%
\left( rR\right) ^{n-1}},  \nonumber  \label{Ftr} \\
F_{\phi _{i}r} &=&-\frac{a_{i}}{\Xi }F_{tr}.
\end{eqnarray}%
where $q$, an integration constant, is the charge parameter of the
black brane. Using metric (\ref{Met3}) and the Maxwell fields
(\ref{Ftr}), one can show that equations (\ref{FE1})-(\ref{FE2})
have solutions of the form
\begin{eqnarray}
&&U(r)=-\left( \frac{c}{r}\right) ^{n-2}\left[ 1-\left( \frac{b}{r}\right)
^{n-2}\right] ^{1-\gamma \left( n-2\right) }-\frac{2\Lambda r^{2}}{n(n-1)}%
\left[ 1-\left( \frac{b}{r}\right) ^{n-2}\right] ^{\gamma },  \label{U} \\
&&W(r)=\Bigg{\{}-\left( \frac{c}{r}\right) ^{n-2}\left[ 1-\left( \frac{b}{r}%
\right) ^{n-2}\right] ^{1-\gamma \left( n-2\right) }-\frac{2\Lambda r^{2}}{%
n(n-1)}\left[ 1-\left( \frac{b}{r}\right) ^{n-2}\right] ^{\gamma }\Bigg
{\}}  \nonumber \\
&&\times \left[ 1-\left( \frac{b}{r}\right) ^{n-2}\right] ^{\gamma (n-3)},
\label{W} \\
&&\Phi (r)=\frac{n-1}{4}\sqrt{\gamma (2+2\gamma -n\gamma )}\ln \left[
1-\left( \frac{b}{r}\right) ^{n-2}\right] ,  \label{Phi} \\
&&R(r)=\left[ 1-\left( \frac{b}{r}\right) ^{n-2}\right] ^{\gamma /2},
\label{R}
\end{eqnarray}%
Here $b$ and $c$ are integration constants and the constant $\gamma $ is
\begin{equation}
\gamma =\frac{2\alpha ^{2}}{(n-2)(n-2+\alpha ^{2})}.  \label{gamma}
\end{equation}%
The charge parameter $q$ is related to $b$ and $c$ by
\begin{equation}
q^{2}=\frac{(n-1)(n-2)^{2}}{2(n-2+\alpha ^{2})}b^{n-2}c^{n-2}.  \label{Q}
\end{equation}%
When ($\alpha =0=\gamma $), the above solution recovers the asymptotically
AdS charged rotating black branes presented in \cite{Deh3,awad}. In the
particular case $n=3$ these solutions reduce to the asymptotically AdS
charged rotating dilaton black strings \cite{Sheykhi3}. For $n=3$ and $%
\alpha =0$ they reduce to charged rotating black string solutions presented
in \cite{Lem0}. Inserting solutions (\ref{U})-(\ref{R}) into the Maxwell
fields (\ref{Ftr}), they can be simplified as
\begin{eqnarray}
F_{tr} &=&\frac{q\Xi }{r^{n-1}},  \nonumber  \label{Ftr2} \\
F_{\phi _{i}r} &=&-\frac{a_{i}}{\Xi }F_{tr}.
\end{eqnarray}%
As one can see from Eq. (\ref{Ftr2}), in the background of AdS universe, the
dilaton field does not exert any direct influence on the matter fields $%
F_{tr}$ and $F_{\phi _{i}r}$'s $(i=1,...,k)$, however, the dilaton field
modifies the geometry of the spacetime as it participates in the field
equations. This is in contrast to the solutions presented in \cite{Sheykhi0}%
. The solutions of Ref. \cite{Sheykhi0} are neither asymptotically flat nor
(A)dS and the gauge field crucially depends on the scalar dilaton field. The
gauge potential $A_{\mu }$ corresponding to the electromagnetic tensor (\ref%
{Ftr2}) can be obtained as
\begin{equation}
A_{\mu }=-\frac{q}{(n-2)r^{n-2}}\left( \Xi \delta _{\mu
}^{t}-a_{i}\delta _{\mu }^{i}\right) \hspace{1.4cm}{\text{(no sum
on i)}}.  \label{Pot}
\end{equation}
\begin{figure}[tbp]
\epsfxsize=5cm \centerline{\epsffile{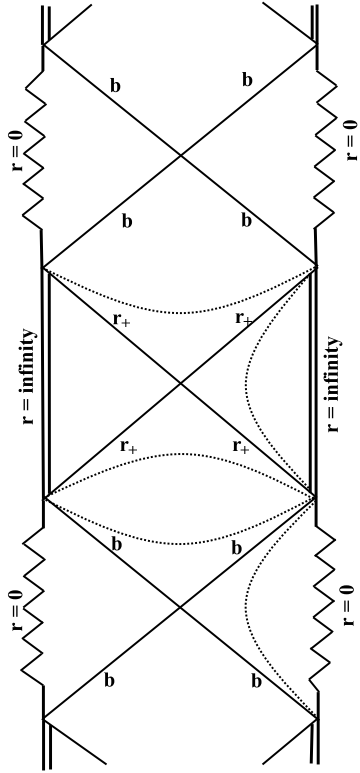}}
\caption{Penrose diagram for black brane with two horizon
($\alpha=0$) located at $r=b$ and $r=r_{+}$. The dotted curves
represent $r=\mathrm{const}.$} \label{Fig1penrose}
\end{figure}
\begin{figure}[tbp]
\epsfxsize=4cm \centerline{\epsffile{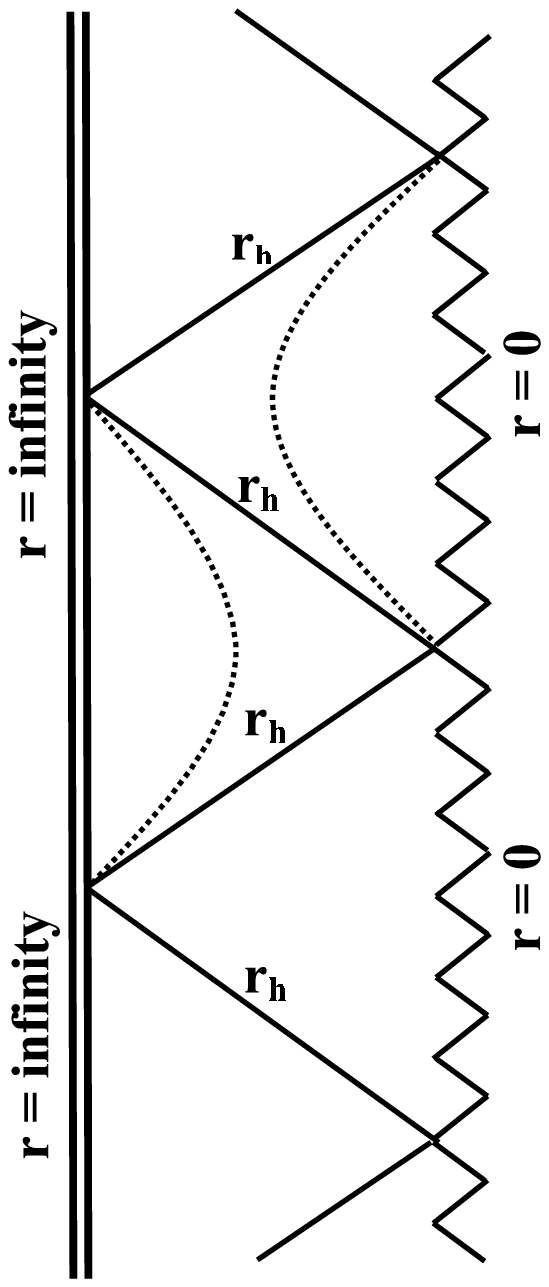}}
\caption{Penrose diagram for extreme black brane with $\alpha=0$
and one horizon located at $r=r_{+}=r_{\mathrm{h}}$. The dotted
curves represent $r=\mathrm{const}.$} \label{Fig2penrose}
\end{figure}
\begin{figure}[tbp]
\epsfxsize=6cm \centerline{\epsffile{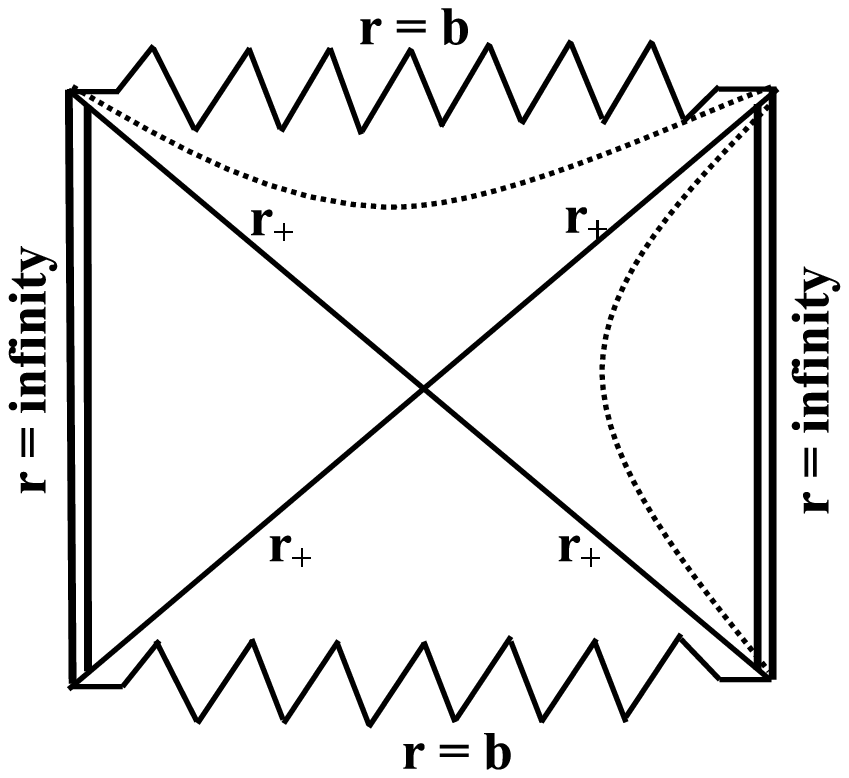}}
\caption{Penrose diagram for black brane with $\alpha \neq 0$ and
one horizon located at $r_{+}$. The dotted curves represent
$r=\mathrm{const}.$} \label{Fig3penrose}
\end{figure}
This spacetime is asymptotically AdS, since the functions $W(r)$
and $U(r)$ behave as $-2\Lambda \left(r^{2}/[n(n-1)]\right)$ as
$r\rightarrow\infty$. Indeed, for large values of $r$, in four
dimensions $(n=3)$ the functions $W(r)$ and $U(r)$ behave as
$-2\Lambda \left(r^{2}/6+p_{1}r+p_{2}\right)$, while in higher
dimensions ($n\geq 4$) they behave like $-2\Lambda
r^{2}/[n(n-1)]$. Here $p_{1}$ and $p_{2}$ are functions of $\alpha
$. This implies that the falloff rate of the solutions in four
dimension is much slower than in higher dimensions. The
Kretschmann invariant $R_{\mu \nu \lambda \kappa }R^{\mu \nu
\lambda \kappa }$ and the Ricci scalar $R$ diverge at $r=0$ and
therefore there is an essential singularity located at $r=0$. For
all $\alpha $, the surface $r=r_{+}$ is an event horizon (the
positive root of Eq. $W(r=r_{+})=0$). The surface $r=b$ is a
curvature singularity except for the case $\alpha =0$ when it is a
nonsingular inner horizon. This is consistent with the idea that
the inner horizon is unstable in the Einstein-Maxwell theory.
Therefore, our solutions describe black branes only in the case
$b<r_{+}$ \cite{Hor3}. For $\alpha=0$ the metric (\ref{metric}) is
real in the range $0 \leq r < \infty$, while for $\alpha > 0$, it
is real only in the range $b \leq r < \infty$. Thus, in order to
have a real metric, we restrict the spacetime to the region $r
\geq b$. We plot Penrose diagrams of spacetime in Figs.
\ref{Fig1penrose}-\ref{Fig3penrose}. From these figures we find
out that the casual structure is asymptotically well behaved. It
is notable to mention that in contrast to the solutions of Ref.
\cite{DPH}, here we have a spacelike singularity with one horizon
and the solutions are asymptotically AdS, at all times, in the
presence of dilaton field.  It is also worthwhile to
note that the dilaton field $\Phi (r)$ and the electromagnetic fields $%
F_{tr}$ and $F_{\phi _{i}r}$'s become zero as $r\rightarrow \infty
$. As in the case of rotating black hole solutions of the Einstein
gravity, the above metric has both Killing and event horizons. The
Killing horizon is a null surface whose null generators are
tangent to a Killing field. It is easy to see that the Killing
vector
\begin{equation}
\chi =\partial _{t}+{{{\sum_{i=1}^{k}}}}\Omega _{i}\partial _{\phi _{i}},
\label{chi}
\end{equation}%
is the null generator of the event horizon, where $k$ denote the number of
rotation parameters \cite{Deh4}. We can obtain the temperature and angular
velocity of the horizon by analytic continuation of the metric. The
analytical continuation of the Lorentzian metric by $t\rightarrow i\tau $
and $a\rightarrow ia$ yields the Euclidean section, whose regularity at $%
r=r_{+}$ requires that we should identify $\tau \sim \tau +\beta _{+}$ and $%
\phi _{i}\sim \phi _{i}+\beta _{+}\Omega _{i}$, where $\beta _{+}$ and $%
\Omega _{i}$ 's are the inverse Hawking temperature and the $i$th component
of angular velocity of the horizon, respectively. The Hawking temperature of
the black brane on the horizon $r_{+}$ can be calculated using the relation
\begin{equation}
T_{+}=\beta _{+}^{-1}=\left( \frac{U^{\text{ }^{\prime }}}{4\pi \Xi \sqrt{U/W%
}}\right) _{r=r_{+}}.
\end{equation}%
where a prime denotes derivative with respect to $r$. It is a matter of
calculation to show that
\begin{eqnarray}
T_{+} &=&\frac{1}{4\pi \Xi r_{+}}\Bigg{\{}n+(n-1)[\gamma (n-2)-2]\left(
\frac{b}{r_{+}}\right) ^{n-2}\Bigg {\}}\left( \frac{c}{r_{+}}\right) ^{n-2}%
\left[ 1-\left( \frac{b}{r_{+}}\right) ^{n-2}\right] ^{-\gamma
(n-1)/2}
\label{Tem} \\
\Omega _{i} &=&\frac{a_{i}}{\Xi l^{2}}.  \label{Om1}
\end{eqnarray}%
where we have used equation $W(r=r_{+})=0$ for omitting $\Lambda
$. It is easy to check that for $\alpha \geq \sqrt{n}$, the
temperature is always positive, while for $\alpha <\sqrt{n}$  the
temperature is positive definite provided we have
\begin{eqnarray}
r_{+}>b\left( \frac{-n}{(n-1)[\gamma (n-2)-2]}\right) ^{-1/(n-2)}
\end{eqnarray}
The entropy of the black brane typically satisfies the so called
area law of the entropy which states that the entropy of the black
hole is a quarter of the event horizon area \cite{Beck}. This near
universal law applies to almost all kinds of black holes,
including dilaton black holes/branes, in Einstein gravity
\cite{hunt}. Denoting the volume of the hypersurface boundary at
constant $t$ and $r$ by $V_{n-1}=(2\pi )^{k}\Sigma _{n-k-1}$, we
can show that the entropy per unit volume $V_{n-1}$ of the black
brane is
\begin{equation}
{S}=\frac{\Xi r_{+}^{n-1}}{4l^{n-2}}\left[ 1-\left(
\frac{b}{r_{+}}\right) ^{n-2}\right] ^{\gamma (n-1)/2}.
\label{Entropy}
\end{equation}%
The mass per unit volume $V_{n-1}$ of the black brane can be
calculated through the use of Eq. (\ref{Mastot}). We find
\begin{eqnarray}
&&{M}=\frac{(3\Xi ^{2}-1)c}{16\pi l}+\frac{\alpha ^{2}(\alpha ^{2}-1)b^{3}}{%
24\pi l^{3}(\alpha ^{2}+1)^{3}}\hspace{0.7cm}\mathrm{for}\ \ n=3,
\label{M4D} \\
&&{M}=\frac{(n\Xi ^{2}-1)c^{n-2}}{16\pi
l^{n-2}}\hspace{3.1cm}\mathrm{for}\ \ n\geq 4.  \label{MnD}
\end{eqnarray}%
Let us note that the mass expression in four dimension differs
from higher dimensions. This is due to the fact that, for large
values of $r$, the falloff rate of the solutions in four and
higher dimensions are different. Thus, the mass in four dimension
depends on the dilaton coupling constant $\alpha $ while in higher
dimensions it is independent of $ \alpha $ and coincides with the
mass of charged rotating black branes in Einstein gravity
\cite{Deh3}. The angular momentum per unit volume $V_{n-1}$ of the
black brane can be calculated through the use of Eqs.
(\ref{Angtot}). We obtain
\begin{equation}
J_{i}=\frac{n\Xi c^{n-2}a_{i}}{16\pi l^{n-2}}.  \label{J}
\end{equation}%
For $a_{i}=0$ ($\Xi =1$), the angular momentum per unit volume vanishes, and
therefore $a_{i}$'s are the rotational parameters of the spacetime. Next, we
calculate the electric charge of the solutions. To determine the electric
field we should consider the projections of the electromagnetic field tensor
on special hypersurfaces. The normal to such hypersurfaces is
\begin{equation}
u^{0}=\frac{1}{N},\text{ \ }u^{r}=0,\text{ \ }u^{i}=-\frac{V^{i}}{N},
\end{equation}%
where $N$ and $V^{i}$ are the lapse function and shift vector. Then the
electric field is $E^{\mu }=g^{\mu \rho }e^{\frac{-4\alpha \phi }{n-1}%
}F_{\rho \nu }u^{\nu }$, and the electric charge per unit volume $V_{n-1}$
can be found by calculating the flux of the electric field at infinity,
yielding
\begin{equation}
{Q}=\frac{\Xi q}{4\pi l^{n-2}}.  \label{chden}
\end{equation}%
The electric potential $U$, measured at infinity with respect to
the horizon, is defined by \cite{Cham2,Cal}
\begin{equation}
U=A_{\mu }\chi ^{\mu }\left\vert _{r\rightarrow \infty }-A_{\mu }\chi ^{\mu
}\right\vert _{r=r_{+}},
\end{equation}%
where $\chi $ is the null generators of the event horizon given by Eq. (\ref%
{chi}). It is a matter of calculation to show that
\begin{equation}
U=\frac{q}{(n-2)\Xi {r_{+}^{n-2}}}.  \label{Pott}
\end{equation}

\section{Conclusions}

\label{Con} It is well-known that in the presence of
Liouville-type dilaton potential, which is regarded as the
generalization of the cosmological constant, the asymptotic
behavior of the solutions change to be neither asymptotically flat
nor (A)dS. As a matter of fact, with the exception of a pure
cosmological constant, no dilaton (A)dS solution exists with the
presence of only one or two Liouville-type potential \cite{MW}. In
this paper, with an appropriate combination of \textit{three}
Liouville-type dilaton potentials, we obtained a new class of
charged rotating black brane solutions in $(n+1)$-dimensional
Einstein-Maxwell-dilaton gravity with cylindrical or toroidal
horizons in the background of AdS spaces and investigated their
properties. We found a suitable counterterm which removes the
divergences of the action in the presence of dilaton potential in
all higher dimensions. We ploted the Penrose diagrams associated
with these spacetimes. These diagrams show that for $\alpha=0$,
the solutions can be interpreted as black brane with two event
horizons, an extreme black brane or a naked singularity provided
the parameters of the solutions are chosen suitably. In this case
we encounter a timelike singularity which is located at $r=0$. We
found out that for $\alpha > 0$, the solutions represent black
brane with one horizon. In this case we have a spacelike
singularity at $r=b$. We also computed the conserved and
thermodynamic quantities of the black branes by using the
conterterm method inspired by AdS/CFT correspondence.
Interestingly enough, we found that the mass expression in four
dimension depends on the dilaton coupling constant $\alpha $,
while in higher dimensions it is independent of $\alpha$. This can
be understood easily, since in four dimension the functions $W(r)$
and $U(r)$ behave as $-2\Lambda \left(r^{2}/6+p_{1}r+p_{2}\right)$
for large $r$, while in higher dimensions ($n\geq4$) they behave
like $-2\Lambda r^{2}/[n(n-1)]$.

\acknowledgments{We thank the anonymous referee for constructive
comments. We are also grateful to Prof. M.H. Dehghani for helpful
discussions. This work has been supported by Research Institute
for Astronomy and Astrophysics of Maragha.}

\end{document}